\documentstyle[aps,pre,psfig]{revtex}

\begin{document}
\draft
\twocolumn[\hsize\textwidth\columnwidth\hsize\csname @twocolumnfalse\endcsname

\title{Comment on ``$1/f$ noise in the Bak-Sneppen model''}
\author{J\"orn Davidsen \cite{byline} and Norbert L\"uthje \cite{byline2}}
\address{Institut f\"ur Theoretische Physik und Astrophysik, Christian-Albrechts-Universit\"at,\\
Olshausenstra\ss e 40, 24118 Kiel, Germany}
\date{August 18, 2000}
\maketitle

\begin{abstract}
Contrary to the recently published results by Daerden and Vanderzande (Phys. Rev. E {\bf 53}, 4723 (1996)), we show that the time correlation function in the random neighbor version of the Bak-Sneppen model can be well approximated by an exponential giving rise to a $1/f^2$ power spectrum.
\end{abstract}
]

\narrowtext

Recently, an exact solution of the random neighbor version of the Bak-Sneppen model was presented by de Boer and co-workers \cite{boe}. They derived a master equation for the probability $P_n(t)$ that $n$ of out of $N$ numbers have a value less than a fixed value $\lambda$ at (discrete) time $t$. In the limit $N \to \infty$ and $\lambda = 1/2$, $P_n$ has the scaling form $P_n(t) = \frac{1}{\sqrt{N}} f(x=n/\sqrt{N}, \tau=t/N)$. Inserting this expression into the master equation gives the following Fokker-Planck equation for $f(x,\tau)$ with a reflecting boundary at $x=0$:
\begin{eqnarray}
\frac{\partial f}{\partial \tau} = \frac{1}{4} \; \frac{\partial^2 f}{\partial x^2} + \frac{\partial}{\partial x} (xf).
\end{eqnarray}
Consequently, the random neighbor version of the Bak-Sneppen model for $N \to \infty$ is just an Ornstein-Uhlenbeck process, i.e., Brownian motion in a parabolic potential. Given the initial condition $f(x,0) = \delta (x-y)$, the solution is
\begin{eqnarray}
f(x,\tau) = \sqrt{\frac{2}{\pi (1 - \exp^{-2 \tau})}} \exp^{\frac{-2 (x - y \exp^{- \tau})^2}{1 - \exp^{-2 \tau}}}. 
\end{eqnarray}

\begin{figure}
\centerline{\psfig{figure=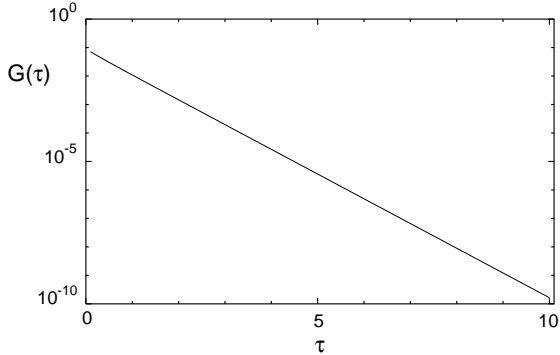,width=\columnwidth,clip=}}
\caption{Plot of the correlation function given by Eq. \ref{cor}.}
\label{pic1}
\end{figure}

\begin{figure}
\centerline{\psfig{figure=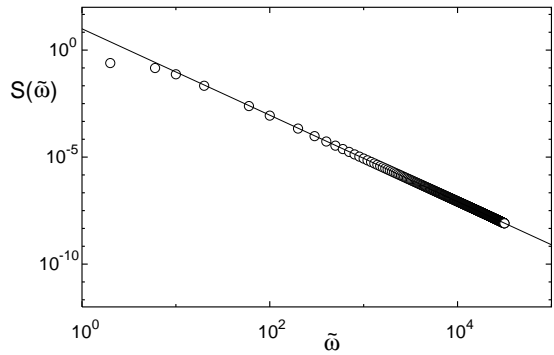,width=\columnwidth,clip=}}
\caption{Plot of the power spectrum given by a numerical Fourier transformation of Eq. \ref{cor}. The solid line with exponent $-2$ is drawn for reference.}
\label{pic2}
\end{figure}

It follows for the autocorrelation function $G(\tau)$ of the time signal $x(\tau)$
\begin{eqnarray}
G(\tau) = \frac{1}{4} \exp^{- a |\tau|},
\end{eqnarray}
where $a \equiv 1$. This directly gives the power spectral density $S(\tilde\omega)$ via a Fourier transform of $G(\tau)$
\begin{eqnarray}
S(\tilde\omega) = \frac{1}{2} \frac{a}{a^2 + \tilde\omega^2}.
\label{pow}
\end{eqnarray}
Going back to the unscaled variables leads to
\begin{eqnarray}
S_P(\omega) = \frac{1}{2} \frac{a}{\frac{a^2}{N^2} + \omega^2},
\end{eqnarray}
which is only valid for low frequencies. Hence, the power spectral density of the signal $n(t)$ decays as $1/f^2$ and for very low frequencies it even becomes constant. However, the above calculation was carried out without applying the boundary condition at $x = 0$. Nevertheless, it is already clear from a physical point of view that the functional form of $G(\tau)$ will not change drasticly by incorporating a reflecting boundary. This is supported mathematically by the fact that one simply has to use the method of images. This was done in Ref. \cite{dae} giving
\begin{eqnarray}
G(\tau) &=& \frac{1}{8 \pi} [1 - \exp^{-2 \tau}]^{3/2} [F(1,2,3/2,r_-(\tau))\nonumber\\
&& + F(1,2,3/2,r_+(\tau)) - F(1,2,5/2,r_-(\tau))/3\nonumber\\
&& - F(1,2,5/2,r_+(\tau))/3] - \frac{1}{2 \pi},
\label{cor}
\end{eqnarray}
where $F(a,b,c,z)$ is the hypergeometric function and where $r_\pm(\tau) = \frac{1}{2} [1 \pm \exp(-\tau)]$. In Fig. \ref{pic1}, $G(\tau)$ is shown. We clearly find an exponential behavior with $a = 0.869 \pm 0.008$ for $0.1 < \tau < 10$ giving rise to a power spectral density as in Eq. (\ref{pow}). This is confirmed by a numerical Fourier transform of Eq. (\ref{cor}) (see Fig. \ref{pic2}). Here, it has to be noted that $G(\tau)$ is an even function, i.e., $G(\tau) = G(-\tau)$. This also ensures that the Fourier transform $S(\tilde\omega)$ is real.\\
In the case of the Bak-Sneppen model with one next neighbor, we also cannot confirm the results presented in Ref. \cite{dae}. A direct simulation of the time signal gives $S_P(\omega) \propto 1/\omega^{1.5}$ over 2 decades for a system size of $N = 8192$. Hence, although the power spectral density in the Bak-Sneppen model decays as a power law, the exponent is far from one. This is also true for a different definition of the time signal \cite{mas}.\\
In conclusion, there is no sign of $1/f$ noise in the random neighbor version of the Bak-Sneppen model and even in the next neighbor version there is no $1/f$ noise in the strict sense.

\acknowledgments

J. Davidsen would like to thank the Land Schleswig-Holstein, Germany, for the financial support granted to him.


\end{document}